\documentstyle[preprint,prl,aps]{revtex}

\input epsf

\begin{document}
\title{Magnetic Field Dependence of Ultracold Inelastic Collisions near a Feshbach Resonance}
\author{J. L. Roberts, N. R. Claussen, S. L. 
Cornish, and C. E. Wieman}
\address{JILA, National Institute of Standards and Technology and the University of Colorado, and the Department of Physics, University of Colorado, Boulder, Colorado\\ 80309-0440}
\date{\today}
\maketitle

\begin{abstract}
Inelastic collision rates for ultracold $^{85}$Rb atoms in the F=2, m$_{f}$=--2 state have been measured as a function of magnetic field.  At 250 Gauss (G), the two- and three-body loss rates were measured to be K$_{2}$=1.87$ \pm$0.95$ \pm$0.19 x 10$^{-14}$ cm$^{3}$/s and K$_{3}$=4.24$^{+0.70}_{-0.29} \pm$0.85 x 10$^{-25}$ cm$^{6}$/s respectively.  As the magnetic field is decreased from 250 G towards a Feshbach resonance at 155 G, the inelastic rates decrease to a minimum and then increase dramatically, peaking at the Feshbach resonance.  Both two- and three-body losses are important, and individual contributions have been compared with theory.
 
\end{abstract}


Feshbach resonances have recently been observed in a variety of cold atom interactions, including elastic scattering \cite{roberts98}, radiative collisions and enhanced inelastic loss \cite{vuletic99}, photoassociation \cite{courteille98}, and the loss of atoms from a Bose-Einstein condensate (BEC) \cite{inouye98}.  By changing the magnetic field through the resonance, elastic collision rates can be changed by orders of magnitude and even the sign of the atom-atom interaction can be reversed \cite{tiesinga9293}.  The work in Ref. \cite{inouye98} has received particular attention because the BEC loss rates were extraordinarily high and several proposals for exotic coherent loss processes have been put forward \cite{timmermans99,abeelen99,yurovsky99}.  However, even ordinary dipole relaxation and three-body recombination are expected to show dramatic enhancements by the Feshbach resonance \cite{nielsen99,esry99,burke98,williamsprivate,moerdijk96,fedichev96}, and the calculations of these ordinary enhancements have never been tested.  This has left many outstanding questions as to the nature of inelastic losses near a Feshbach resonance.  How large are the dipole relaxation and three-body recombination near the Feshbach resonance and how accurate are the calculations of these quantities?  How much of the observed condensate losses in Ref. \cite{inouye98} are due to these more traditional mechanisms and how much arise from processes unique to condensates?  How severe are the``severe limitations'' \cite{inouye98} that inelastic loss puts on the use of Feshbach resonances to change the s-wave scattering length in a BEC?  The nature of the inelastic losses near Feshbach resonances also has important implications for efforts to create BEC in $^{85}$Rb, because these losses play a critical role in determining the success or failure of evaporative cooling.  Thus it is imperative to better understand the nature of inelastic collisions between ground state atoms near a Feshbach resonance.

In this paper we present the study of the losses of very cold $^{85}$Rb atoms from a magnetic trap as a function of density and magnetic field.  There are two types of inelastic collisions that induce loss from a magnetic trap.  The first is dipolar relaxation where two atoms collide and change spin states.  The second process is three-body recombination, where three atoms collide and two of those atoms form a molecule.  Measuring the losses as a function of density has allowed us to determine the two-body and three-body inelastic collision rates, while measuring the variation of these losses as a function of magnetic field 
has allowed us to find out how the Feshbach resonance affects them.  In contrast to the work of Ref. \cite{inouye98} we observe a pronounced dip in the inelastic losses near the Feshbach resonance.  This will make it possible to create $^{85}$Rb BECs with a positive scattering length.

To study these losses we needed a cold, dense $^{85}$Rb sample.  This was obtained through evaporative cooling with a double magneto-optic trap (MOT) system as described in Ref. \cite{myatt96}.  The first MOT repeatedly collected atoms from a background vapor and those atoms were transferred to another MOT in a low-pressure chamber.  Once the desired load size was achieved, the MOTs were turned off and a baseball-type Ioffe-Pritchard magnetic trap was turned on, resulting in a trapped atom sample of about 3x10$^{8}$ F=2, m$_{f}$=--2 $^{85}$Rb atoms at 45 $\mu K$.  Forced radio-frequency (rf) evaporation was used to increase the density of the atoms while decreasing the temperature.  Because of the high ratio of inelastic/elastic collision rates in $^{85}$Rb and the dependence of the ratio on magnetic field ($B$) and temperature ($T$), trap conditions must be carefully chosen to achieve efficient cooling. We evaporated to the desired temperature and density (typically 3x10$^{11}$ cm$^{-3}$ and 500-700 $nK$) at a final field of $B$=162 Gauss (G) \cite{fieldchoice}.

We then adiabatically changed the DC magnetic field to various values and measured the density of the trapped atom cloud as a function of time.  We observed the clouds with both nondestructive polarization-rotation imaging \cite{bradley97} and destructive absorption imaging.  In both cases, the cloud was imaged onto a CCD array to determine the spatial size and number.  The nondestructive method allowed us to observe the time evolution of the number and spatial size of a single sample.  The destructive method required us to prepare many samples and observe them after different delay times, but had the advantages of better signal to noise for a single image and larger dynamic range.  A set of nondestructive imaging data is shown in Fig. 1.  In addition to these techniques, we made a redundant check of the number by recapturing the atoms in the MOT and measuring their fluorescence \cite{townsend95}.

We measured the time-dependence of the number of atoms and the spatial size, and from these measurements determined the density.  We assigned a 10\% systematic error to our density determination, based primarily on the error in our measurement of number.  The value of $B$ was measured in the same way as in Ref. \cite{roberts98}:  the rf frequency at which the atoms in the center of the trap were resonantly spin-flipped was measured and the Breit-Rabi equation was then used to determine the magnitude of $B$.  The magnetic field width of the clouds scaled as $T^{1/2}$ and was 0.39 G FWHM at 500 $nK$.

In each data set we observed the evolution of the sample while a significant fraction (20--35\%) of the atoms was lost.  The temperature of the sample also increased as a function of time.  This heating rate scaled with the inelastic rates and therefore varied with $B$.  Since the volume and temperature in a magnetic trap are directly related, the change in the volume (typically 50\%) with time was fit to a polynomial---usually a straight line was sufficient within our precision.  The number (N) as a function of time was then fit to the sum of three-body and two-body loss contributions given by 
\begin{equation}
\dot{N}=-K_{2} \langle n \rangle N - K_{3} 
\langle n^{2} \rangle N - \frac{N}{\tau}.
\end{equation}
Here $\langle n \rangle = \frac{1}{N} \int n^{2}({\bf x}) d^{3}x$ is the density-weighted density and $\langle n^{2} \rangle = \frac{1}{N} \int n^{3}({\bf x}) d^{3}x$.  The time evolution of the volume is contained in the computation of the density as a function of time.  The background loss rate is given by $\tau$.  It was independently determined by looking at low-density clouds for very long times.  Typically, it was 450 seconds, with some slight dependence on $B$ ($<$20\%), particularly near the Feshbach resonance.

The change in the combined inelastic rates as a function of $B$ around the Feshbach resonance is shown in Fig. 2(a).  All of the inelastic data shown in Fig. 2(a) were taken with initial temperatures near 600 $nK$.  The points less than 157 G were taken with initial densities within 10\% of 1x10$^{11}$cm$^{-3}$, while the points higher than 157 G were taken with initial densities that were ~2.7x10$^{11}$cm$^{-3}$.  The decrease in initial density below 157 G was due to ramping through the high inelastic loss region after forming the sample at 162 G.  Because of this density variation, we have used a different weighting, parameter $\beta$, in the sum of the two- and three-body rate constants for the two regions in Fig. 2(a).

Also shown in Fig. 2(a) is the elastic rate determined previously \cite{roberts98}.  This elastic rate was measured by forming atom samples similar to the ones in this work (but at much lower density), forcing them out of thermal equilibrium, and then observing their reequilibration time.  The shape of the inelastic rate vs. $B$ roughly follows that of the elastic rate vs. $B$.  In particular, the peak of the inelastic rate occurs at 155.4$\pm$0.5 G, identical to the position of the elastic peak at 155.2$\pm$0.4 G within the error. Also, just as is the case for the elastic rates, the inelastic loss rates around the Feshbach resonance vary by orders of magnitude.  The peak in the inelastic rate is much less symmetric, however.  Another interesting feature is that the loss rates not only increase near the elastic rate peak but decrease near its minimum.  The field where the loss is minimum, 173.8$\pm$2.5 G, is higher than the minimum of the elastic rate at 166.8$\pm$0.3 G.

Even though the two- and three-body inelastic rates have different density dependencies, it is difficult to separate them.  Figure 1 shows how a purely two-body or purely three-body loss curve fits equally well to a typical data set.  An excellent signal-to-noise ratio or, equivalently, data from a large range of density are required to determine whether the loss is three-body, two-body, or a mixture of the two.  To better determine density dependence, we decreased the initial density by up to a factor of 10 for several $B$ values.  For fields with relatively high loss rates, the signal-to-noise ratio was adequate to distinguish between two- and three-body loss.  However, where the rates were lower this was not the case.

Figure 2(b) shows the two-body inelastic rate determined from the density-varied data and Fig. 2(c) shows the same for the three-body rate.  The character of the inelastic loss clearly changes as one goes from higher to lower field (right to left in Fig. 2).  On the far right of the graph, at $B$ = 250 G, the inelastic rate is dominated by a three-body process.  From $B$ = 250 G down to $B$ = 174 G the inelastic rate decreases to a minimum and then begins to increase again.  Near the minimum, we could not determine the loss character, but at 162 G the losses are dominated by a two-body process at this density.  At 158 G the two-body process is still dominant and rising rapidly as one goes toward lower field.  However, by $\sim$157 G it has been overtaken by the three-body recombination that is the dominant process at 157-145 G.  At lower fields, both two-body and three-body rates contribute significantly to the total loss at these densities.

Along with varying the density, the initial temperature was varied at $B$ = 145,156,160, and 250 G.  There was no significant rate change in the loss rate for temperatures between 400 and 1000 $nK$ for the 160 and 250 G points.  The combined loss rate increased by a factor of 8 at 156 G from 1 $\mu K$ to 400 $nK$, and by a factor of 2 at 145 G.  This temperature dependence near the peak is expected in analogy to the temperature dependence of the elastic rates \cite{unitarity}. The fact that the loss rates near the peak exhibit both a two- and three-body character plus a temperature dependence and considerable heating make interpreting the data challenging.  This introduces additional uncertainty that is included in the error bars in Fig. 2(a-c).

A calculated dipolar loss rate is compared with the data in Fig. 2(b).  This was calculated by finding the S-matrix between the trapped and untrapped states using carefully determined Rb-Rb molecular potentials \cite{williamsprivate}.  Keeping in mind the significance of error bars on a log plot, the agreement is reasonably good.  In the region between $B$ = 160 and 167 G where the determination of K$_{2}$ is not complicated by three-body loss and temperature dependence, the agreement is particularly good.

Likewise, we show a prediction of the three-body recombination rate in Fig. 2(c) \cite{esry99}.  It is predicted that the recombination rate scales as the s-wave scattering length to the fourth power (a$_{s}^{4}$) for both positive and negative a$_{s}$, although with a smaller coefficient for the positive case \cite{nielsen99,esry99,fedichev96}.  Since a$_{s}$ varies across the Feshbach resonance (see Fig. 2(a), upper plot), the recombination rate is expected to change \cite{scatlength}.  Temperature dependence has not been included in the prediction.  Qualitatively, the main features of the predicted three-body recombination match the data.  The three-body rate decreases as a$_{s}$ decreases, and it increases rapidly where a$_{s}$ diverges at $B$ = 155 G.  From 155 G to 167 G, a$_{s}$ is positive and the three-body loss is much smaller than it is at $B$ fields with a comparable negative a$_{s}$, as expected from theory.  This overall level of agreement is reasonable given the difficulties and approximations in calculating three-body rates.

The marked dependence on $B$ of both the elastic and inelastic rates has important implications for the optimization of evaporative cooling to achieve BEC in $^{85}$Rb.  As the density increases with evaporative cooling, the absolute loss rate, which is already much greater than in $^{87}$Rb and Na, becomes larger.  However, the important ratio of elastic rate to inelastic rate is both temperature and density dependent and magnetic field dependent.   The fact that the losses are dramatically lower at fields above the Feshbach resonance suggests that it should be possible to devise an evaporation path that will lead to the creation of an $^{85}$Rb BEC \cite{BEC}.

The $^{85}$Rb Feshbach resonance has a profound effect on the two- and three-body inelastic rates, changing them by orders of magnitude.  Far from the resonance at 250 G, the two-body and three-body loss rates are measured to be K$_{2}$ = 1.87$ \pm$0.95$ \pm$0.19 x 10$^{-14}$ cm$^{3}$/s and K$_{3}$=4.24$^{+0.70}_{-0.29} \pm$0.85 x 10$^{-25}$ cm$^{6}$/s respectively.  Here the first error is the statistical error and the second in the systematic uncertainty due to the number determination.  Near the resonance, the two- and three-body rates change by orders of magnitude.  The dependence of the inelastic rates on magnetic field is similar in structure to the dependence of the elastic rate:  the maxima in both rates occur at the same field, while the minima are close but do not coincide.  The total loss is a complicated mixture of both two- and three-body loss processes.  They have different dependencies on field so both have field regions in which they dominate.

We are pleased to acknowledge useful discussions and with Eric Cornell, Jim Burke, Jr., Chris H. Greene, and Carl Williams.  We also thank the latter three and their colleagues for providing us with loss predictions.  This research has been supported by the NSF and ONR.  One of us (S. L. Cornish) acknowledges the support of a Lindemann fellowship.

\begin{figure}
{\caption {Number of atoms versus time ($\bullet$) taken by polarization-rotation imaging at 159 G.  Fits to both purely two-body (dashed line) and purely three-body (solid line) inelastic rates are shown, illustrating the difficulty in separating two- and three-body loss processes.}}
\end{figure}

\begin{figure}
{\caption{(a) Elastic rate (upper plot and $\circ$ symbol) and s-wave scattering length a$_{s}$ (upper plot and solid line) from Ref. [1], and total loss rates (lower plot) vs. magnetic field. The total loss rate is expressed as a sum of the two- and three-body loss as K$_{2} + \beta $K$_{3}$.  Due to initial density differences caused by ramping across the peak, $\beta$ = 1.6 x 10$^{11}$cm$^{-3}$ for the points below $B$=157 G (filled circles) and 4.0 x 10$^{11}$cm$^{-3}$ for the points above (filled triangles).  The vertical lines represent the positions of the elastic rate maximum and minimum at $B$=155 G and $B$=167 G respectively.  (b) The determination of the two-body inelastic rate for several $B$ fields.  The theory prediction from Ref. [12] is shown as a solid line.  The open circles ($\circ$) are the two-body rates determined from the total loss from Fig. 2(a) by assuming (not explicitly measuring) that the loss between 162 G and 166 G is predominantly two-body.  These points are added to aid in comparison with theory.  (c) The determination of the three-body inelastic rate.  The points with down arrows ($\frac{ }{\downarrow}$) are to be interpreted as upper limits on the three-body rate.  The solid line here is a prediction of the loss rate from Ref. [10].  The open circles in Fig. 2(c) are similar to those in 2(b), only in 2(c) we assume (but again do not measure) that the loss above 175 G is predominantly three-body.  The error bars on the 250 G point are relatively small because a large amount of data was taken there.  In addition to the statistical errors shown in Fig. 2(a-c), there is another 10\% systematic uncertainty in K$_{2}$ and 20\% in K$_{3}$ due to the estimated error in our density determination.}}
\end{figure}


\begin{references}
\bibitem{fieldchoice} The choice of 162 G for the final stage of evaporation was made somewhat +arbitrarily.
\end{references}
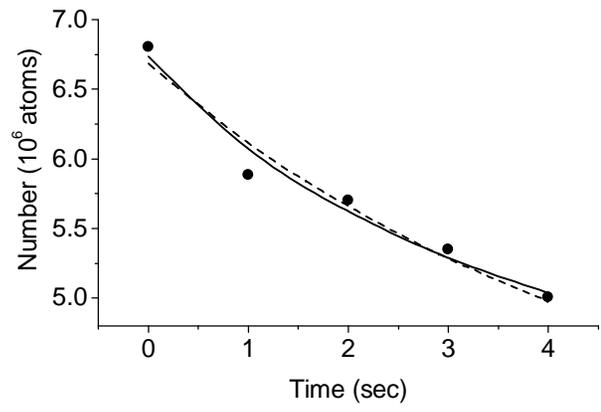
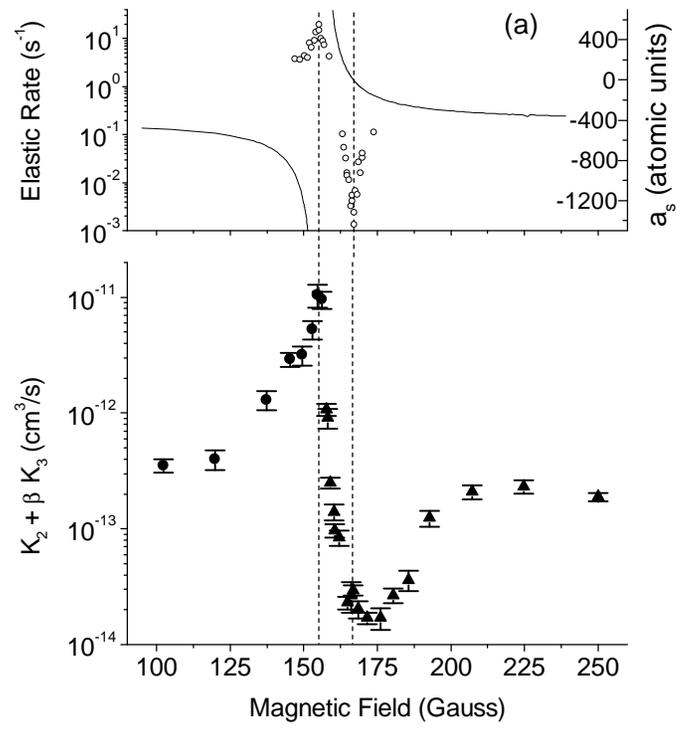
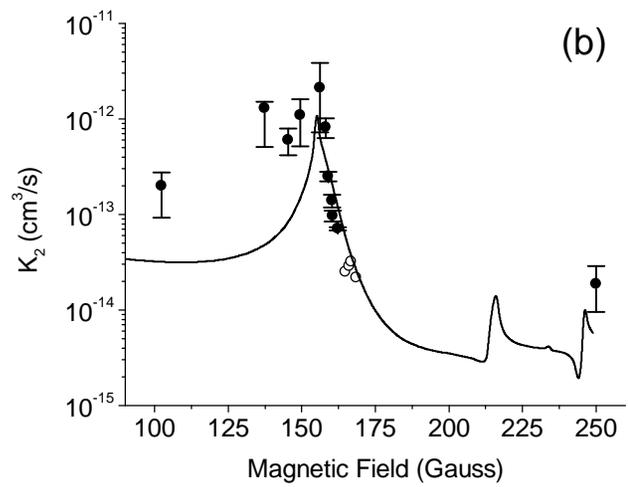
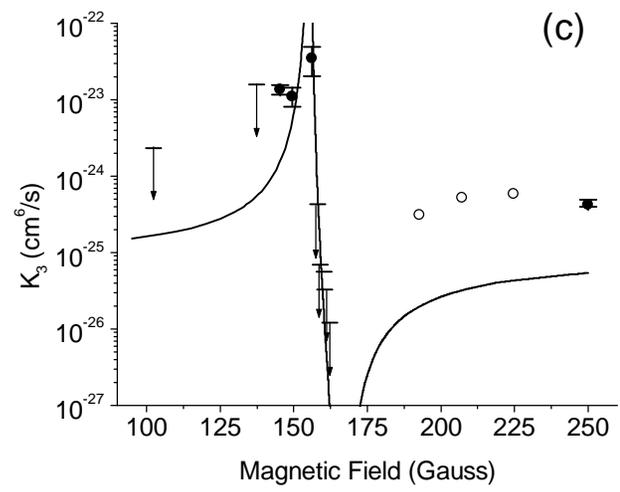
\end{document}